\documentclass[12pt]{iopart}
\begin{document}
\title[]{A Section-Based Queueing-Theoretical Traffic Model for Congestion
and Travel Time Analysis in Networks}
\author{Dirk Helbing} 
\address{Institute of Economics and Traffic, Dresden University of Technology,
Andreas-Schubert-Str. 23, 01062 Dresden, Germany, helbing@trafficforum.org} 
\begin{abstract}
While many classical traffic models treat the spatial extension of streets continuously or
by discretization into cells of a certain length $\Delta x$, 
we will subdivide roads into comparatively long homogeneous road sections of constant capacity
with an inhomogeneity at the end. 
The related model is simple and numerically efficient. 
It is inspired by models of dynamic queueing networks and takes into account 
essential features of traffic flows. Instead of treating single vehicles or velocity profiles,
it focusses on flows at specific cross sections 
and average travel times of vehicles.
\end{abstract}
\pacs{89.40.-a,
47.40.Nm,
47.55.-t
}
\maketitle

In the last decade, physicists have made many significant contributions to traffic modeling
\cite{review}. The modeling approaches reach from cellular automata \cite{CA} 
over microscopic car-following models \cite{follow}
up to macroscopic or fluid-dynamic traffic models \cite{LW,fluid,macro}. 
They have, for example, been the basis for understanding the various breakdown phenomena of traffic 
flow in terms of non-equilibrium phase transitions \cite{phase}.
\par
Classical traffic models are mostly based on the treatment of interacting vehicles, their statistical distribution,
or their density and average velocity as a function of space and time. For numerical reasons, space and
time are often discretized into small intervals $\Delta x$ and $\Delta t$. Depending on the model,
typical values of $\Delta x$ are between 5 and 500 meters. Although drivers are normally interested in
travel times, most models focus on the velocities (from which the travel times are sometimes derived). 
An alternative approach are queueing models (see \cite{qu,Cremer,Wagner,Kern} and references
therein): For example, Cremer and Landenfeld have
developed a microscopic model for saturated urban road networks which is discrete in time.
It distinguishes only vehicles moving with the maximum allowed speed $V_i^0 = 50$km/h
and standing vehicles with the space requirement 6.7m, assuming discharge rates of $Q_{\rm out} = 1/2$s and
parallel forward motion of all queued vehicles when a site becomes empty \cite{Cremer}. The model by
Eissfeldt, Gr\"afe, and Wagner 
eliminates vehicle motion and translates macroscopic flows into microscopic time headways. 
It is based on a section travel time and additional
traffic-state dependent waiting times of cars, taking into account the finite velocity of the upstream
propagation of emptied road space \cite{Wagner}. It can reproduce empirical flows quite
well \cite{Wagner}. In contrast to these microscopic queueing models, Kerner proposes
a macroscopic approach to city traffic relating to the continuity equation \cite{Kern}. It 
assumes a constant density of saturated (queued) traffic and a time-dependent, but
space-independent density of unsaturated traffic. Therefore, it neglects the spatio-temporal propagation of
density variations. 
\par
In the following, we will propose a model which exactly integrates the Lighthill-Whitham model
over finite road sections. Although it formally eliminates the traffic dynamics on the road sections by expressing
it through the dynamics at a few cross sections of the road network, it takes into account continuous (and 
discontinuous) changes in the density of free or congested traffic and their characteristic finite propagation velocities.
The model was inspired by a dynamic model of supply and production networks \cite{queue}.
When we now specify road traffic as a queueing system, we will take into account essential 
traffic characteristics such as the flow-density relation or the properties of extended congestion patterns 
at bottlenecks. In fact, traffic congestion is usually triggered by spatial
inhomogeneities of the road network \cite{phase}, 
and queueing effects are normally not observed along sections of low
capacity, but upstream of the beginning of a bottleneck. Therefore, 
we will subdivide roads into sections $i$ of homogeneous capacity and length $L_i$,
which start at place $x_i$ and
end with some kind of inhomogeneity (i.e. an increase or decrease of capacity) at place $x_{i+1}= x_i+L_i$.
In other words, the end of a road section $i$ is, for example, determined by the location
of an on- or off-ramp, a change in the number of lanes, or the beginning or end of a gradient.
\par
We will derive our model from the continuity equation 
\begin{equation}
 \frac{\partial \rho(x,t)}{\partial t} + \frac{\partial Q_i(x,t)}{\partial x} = 
 \mbox{Source Terms} 
\end{equation}
describing the conservation of the number of vehicles.
Here, $\rho(x,t)$ denotes the vehicle density per lane at place $x$ and time $t$, and
$Q_i(x,t)$ the traffic flow per lane. The source terms originate from ramp flows
$Q_i^{\rm rmp}(t)$, which enter the road at place $x_{i+1}$. Let us define the
arrival rate at the upstream end of road section $i$ by $Q_i^{\rm arr}(t)= I_i Q_i(x_i+dx,t)$,
where $dx$ is a differential space interval and $I_i$ the number of lanes of road
section $i$. Analogously, the departure rate from the downstream end of this section is defined by
$Q_i^{\rm dep}(t) = I_i Q_i(x_{i+1}-dx,t)$. 
The conservation of the number of vehicles implies that the departure rate plus the ramp flow determine
the arrival rate in the next downstream section $i+1$:
\begin{equation}
 Q_{i+1}^{\rm arr}(t) = Q_i^{\rm dep}(t) + Q_i^{\rm ramp}(t) \, .
\label{arr}
\end{equation}
In order to guarantee non-negative flows, we will demand for the ramp flows that the consistency condition
$- Q_i^{\rm dep}(t) \le Q_i^{\rm ramp}(t) \le Q_{i+1}^{\rm arr}(t)$ is always met.
\par
Integrating the continuity equation over $x$ with $x_i < x < x_{i+1}$ provides a conservation
equation for the number $N_i(t)
= \int_{x_i}^{x_{i+1}} dx \; I_i \rho(x,t)$ of vehicles in road section $i$. It changes according to 
\begin{equation}
 \frac{dN_i(t)}{dt} = Q_i^{\rm arr}(t) - Q_i^{\rm dep}(t) = Q_{i-1}^{\rm dep}(t) + Q_{i-1}^{\rm ramp}(t)  
 - Q_{i}^{\rm dep}(t) \, .
\label{length}
\end{equation}
In the following, we will try to express the traffic dynamics and the travel times only through the 
flows at the cross sections $x_i$. For this, we will take into account the following simplified features
of traffic flow: (i) For free flow, i.e. below some critical vehicle density 
$\rho_{\rm cr}$ per lane, the relation between the traffic flow $Q_i$ per lane and the density $\rho$ per lane
can be approximated by an increasing linear relationship, 
while above it, a falling linear relationship 
is consistent with congested flow-density data (in particularly, if the average time gap
$T$ is treated as a time-dependent, fluctuating variable) \cite{gaps}. This implies $Q_i(x,t)
\approx Q_i(\rho(x,t))$ with
\begin{equation}
 Q_i(\rho) = \left\{
\begin{array}{ll}
Q_i^{\rm free}(\rho) = \rho V_i^0 & \mbox{if } \rho < \rho_{\rm cr} \\
Q_i^{\rm cong}(\rho) = ( 1 - \rho / \rho_{\rm jam}) / T & \mbox{otherwise.}  
\end{array}\right. 
\label{Q}
\end{equation}
Here, $V_i^0$ denotes the average free velocity,
$T$ the average time gap, and $\rho_{\rm jam}$ the density per lane
inside of traffic jams. Moreover, we define the free and congested densities by
\begin{equation}
 \rho_i^{\rm free}(Q_i) = Q_i / V_i^0 \quad \mbox{and} \quad
 \rho_i^{\rm cong}(Q_i) = (1-TQ_i)\rho_{\rm jam} \, .
\label{def}
\end{equation}
The quantity $Q_{\rm out} = (1-\rho_{\rm cr}/\rho_{\rm jam})/T$ 
corresponds to the outflow per lane from congested
traffic \cite{outflow}. Depending on the parameter specification,
the model describes a continuous flow-density relation 
(for $\rho_{\rm cr} V_i^0 = Q_{\rm out}$) or a 
capacity drop at the critical density $\rho_{\rm cr}$ and high-flow states immediately before
(if $\rho_{\rm cr} V_i^0 > Q_{\rm out}$). 
(ii) According to shock wave theory \cite{LW}, density variations at place $x$ propagate with velocity
$C(t) = [Q_i(x+dx,t)-Q_i(x-dx,t)]/[\rho(x+dx,t)-\rho(x-dx,t)]$. 
Accordingly, the propagation velocity is $C=V_i^0$ in free traffic, and $C= - c = 
- 1/(T\rho_{\rm max})$ in congested traffic. Therefore, it takes the time period $T_i^{\rm free} =
L_i/V_i^0$ for a perturbation to travel through free traffic, while it takes the time period $T_i^{\rm cong}
= L_i/c$, when the entire road section $i$ is congested.
(iii) Now, remember that congestion in section $i$ starts to form upstream of a
bottleneck, i.e. at place $x_{i+1}$. Let $l_i(t)$ denote the length of the congested area and
$x(t) = x_{i+1}-l_i(t) = x_i+L_i-l_i(t)$ the location of its upstream front. Then, we have free traffic
between $x_i$ and $x_i+L_i-l_i(t)$, i.e. $Q_i(x-dx,t) = Q_i^{\rm arr}(t - (x-x_i)/V_i^0)$ (considering $dx \rightarrow 0$), 
and congested traffic downstream
of $x(t)$, i.e. $Q_i(x+dx,t) = Q_i^{\rm dep}(t - (x_{i+1}-x)/c)$. With $dx/dt = -dl_i/dt = C(t)$ and Eq.~(\ref{def}) we find
\begin{equation}
 \frac{dl_i}{dt} = -\frac{Q_i^{\rm dep}(t-l_i(t)/c)/I_i - Q_i^{\rm arr}(t-[L_i-l_i(t)]/V_i^0)/I_i}
{\rho_i^{\rm cong}(Q_i^{\rm dep}(t-l_i(t)/c)/I_i) - \rho_i^{\rm free}(Q_i^{\rm arr}(t-[L_i-l_i(t)]/V_i^0)/I_i)} \, .
\label{see}
\end{equation}
(iv) The capacity of a {\em congested} road section $i$ is approximated as the outflow 
$Q_{\rm out} = (1-\rho_{\rm cr}/\rho_{\rm jam})/T$ from congested traffic per lane
times the number $I_i$ of lanes, minus the maximum bottleneck strength at the end of this section.
This may be given by an on-ramp flow $Q_i^{\rm ramp}(t) > 0$ or analogously by $(I_{i} - I_{i+1})Q_{\rm out}$ 
in case of a reduction $I_{i+1}-I_i < 0$ in the number of lanes, or in general by some potentially time-dependent value
$\Delta Q_i(t)$ in case of another bottleneck such as a gradient:
\begin{equation}
 Q_i^{\rm cap}(t) = I_i Q_{\rm out} - \max[ Q_i^{\rm ramp}(t), (I_{i}-I_{i+1})Q_{\rm out}, \Delta Q_i(t), 0] \, .
\end{equation} 
Analogously, the maximum capacity $Q_i^{\rm max}(t)$ of the road section $i$ under free flow conditions is
given by the maximum flow $I_i \rho_{\rm cr} V_i^0$ minus the reduction by bottleneck effects:
\begin{equation}
 Q_i^{\rm max}(t) = I_i \rho_{\rm cr} V_i^0 - \max[ Q_i^{\rm ramp}(t), (I_{i}-I_{i+1})\rho_{\rm cr} V_i^0, \Delta Q_i(t), 0] \, .
\end{equation} 
Note that an off-ramp flow $Q_i^{\rm ramp}(t)<0$ does not contribute to these formulas, but
one may reflect the bottleneck effect of weaving flows (i.e. frequent lane changes at exits)
by some $\Delta Q_i(t) \ge 0$. 
(v) We can distinguish three states $S_i(t)$ of road section $i$: $S_i(t) = 0$ corresponds to free traffic,
which assumes that the length $l_i(t)$ of the congested area of road section $i$ is zero and that the 
maximum capacity $Q_i^{\rm max}(t-dt)$ in the last time step $t-dt$, where $dt$ denotes a small time interval,
was not reached by the arrival rate at time $t-dt-T_i^{\rm free}$,  
i.e. $Q_i^{\rm arr}(t-dt-T_i^{\rm free}) < Q_i^{\rm max}(t-dt)$. $S_i(t) = 1$ corresponds
to a completely congested road section $i$, which assumes with $l_i(t) = L_i$ that the congested area expands
over the full section length $L_i$ and that the arrival rate $Q_i^{\rm arr}(t-dt)$ 
in the last time step $t-dt$ was not below
the departure rate at time $t-dt- T_i^{\rm cong}$, 
i.e. $Q_{i}^{\rm dep}(t-dt-T_i^{\rm cong}) \le Q_{i}^{\rm arr}(t-dt)$. Otherwise we have partially
congested traffic in road section $i$ and set $S_i(t) = 2$. Altogether this implies: 
\begin{equation}
 S_i(t) = \left\{
\begin{array}{ll}
0 & \mbox{if $l_i(t) = 0$ and } Q_i^{\rm arr}(t-dt-T_i^{\rm free}) < Q_i^{\rm max}(t-dt), \\
1 & \mbox{if $l_i(t) = L_i$ and } Q_{i}^{\rm dep}(t-dt-T_i^{\rm cong}) \le Q_{i}^{\rm arr}(t-dt), \\
2 & \mbox{otherwise.}
\end{array}\right.
\label{states}
\end{equation}
(vi) Finally, we have to specify the departure rate $Q_i^{\rm dep}(t)$
as a function of the respective traffic situation. 
Focussing on the cross section at location $x_{i+1}$
and considering the directions of information flow (i.e. the propagation direction of density variations),
we can distinguish three different cases:
\begin{itemize}
\item[1.] If we have free traffic in the upstream section $i$ and free or partially congested
traffic in the downstream section $i+1$, density variations propagate downstream and 
the departure rate $Q_i^{\rm dep}(t)$ at time $t$ 
is given as the arrival rate $Q_i^{\rm arr}(t-T_i^{\rm free}) 
= Q_{i-1}^{\rm dep}(t-T_i^{\rm free}) + Q_{i-1}^{\rm ramp}(t-T_i^{\rm free})$,
since the vehicles entering section $i$ at time $t-T_i^{\rm free}$ 
leave the section after an average travel time $T_i$ of $T_i^{\rm free}$.
\item[2.] In the case of partially or completely congested traffic 
upstream and free or partially congested traffic downstream,
the departure rate $Q_i^{\rm dep}(t)$ is given by
the capacity $Q_i^{\rm cap}(t)$ of the congested road section $i$.  
\item[3.] In the case of congested traffic on the entire downstream road section $i+1$,
the departure rate $Q_i^{\rm dep}(t)$ is given by
the departure rate $Q_{i+1}^{\rm dep}(t-T_{i+1}^{\rm cong})$ from the downstream section 
at time $t-T_i^{\rm cong}$ minus the ramp flow $Q_i^{\rm ramp}(t)$ entering at location $x_{i+1}$.
\end{itemize}
Summarizing this, we have
\begin{equation}
 Q_i^{\rm dep}(t) = \left\{
\begin{array}{ll}
\!Q_{i}^{\rm arr}(t-T_i^{\rm free}) & \mbox{$\!\!$if } 
S_{i+1}(t) \ne 1 \mbox{ and } S_i(t) = 0, \\
\!Q_i^{\rm cap}(t) & \mbox{$\!\!$if } S_{i+1}(t) \ne 1 \mbox{ and } S_i(t) \ne 0, \\
\!Q_{i+1}^{\rm dep}(t-T_{i+1}^{\rm cong})-Q_i^{\rm ramp}(t) & \mbox{$\!\!$if } S_{i+1}(t) = 1.
\end{array}
\right.
\label{rates}
\end{equation}
\par
A numerical solution of the above defined section-based queueing-theoretical model is carried out as
follows:
First, determine the states of
the road sections according to Eq.~(\ref{states}). Second, calculate the new arrival and departure 
rates by means of Eqs.~(\ref{arr}) and (\ref{rates}), taking into account
the boundary conditions for the flows at the open ends of the road network. 
Third, if $S_i(t) = 0$, then set $l_i(t+dt) = l_i(t) = 0$.
If $S_i(t) = 1$, then set $l_i(t+dt) = l_i(t) = L_i$.
Otherwise, if $S_i(t) = 2$, determine the length $l_i(t+dt) = l_i(t) + dt \; dl_i(t)/dt$ 
of the congested area in road section $i$ with formula~(\ref{see}). 
Next, continue with the first step for the new time point $t+dt$. It is obvious, that
this numerical solution is significantly more simple and robust than the numerical solution of
the Lighthill-Whitham model, as shock waves (i.e. the interfaces between free and congested traffic)
are treated analytically. 
\par
Let us now derive a general relationship for the average travel 
time $T_i(t)$ of a vehicle that enters road section $i$ at time $t$.
The average travel time $T_i(t)$ is given by the fact that this vehicle will leave section $i$, when (on average)
the $N_i(t)$ vehicles that are in section $i$ at time $t$ have passed the downstream end of that section.
Since the number of vehicles passing the downstream end of section $i$ is given as the time-integral
over the departure rate $Q_i^{\rm dep}(t)$, we have the implicit relationship
\begin{equation}
 N_i(t) = \int\limits_t^{t+T_i(t)} \!\! dt' \, Q_i^{\rm dep}(t') 
=  \int\limits_{-\infty}^{t+T_i(t)} \!\! dt' \, Q_i^{\rm dep}(t') 
 - \int\limits_{-\infty}^{t} \!\! dt' \, Q_i^{\rm dep}(t') \, .
\end{equation}
Identifying Eq.~(\ref{length}) with the time-derivative of this equation 
finally leads to the delay-differential equation
\begin{equation}
 \frac{dT_i(t)}{dt} = \frac{Q_i^{\rm arr}(t)}{Q_i^{\rm dep}(t+T_i(t))} - 1 
 = \frac{Q_{i-1}^{\rm dep}(t) + Q_{i-1}^{\rm ramp}(t)}{Q_i^{\rm dep}(t+T_i(t))} - 1 \, .
\label{travtime}
\end{equation}
According to this, the travel time $T_i(t)$ increases with time, when the arrival rate $Q_i^{\rm arr}$ 
at the time $t$ of entry exceeds the departure rate $Q_i^{\rm dep}$ at the leaving time $t+T_i(t)$, while it decreases when it
is lower. It is remarkable that this formula does not explicitly depend on the velocities on the road section, but
only on the arrival and departure rates. 
The calculation of the
travel time based on the velocity $v(t)$ is considerably more complicated as it
depends on the traffic density $\rho(x(t),t)$ 
at its respective location $x(t)=x_i + \int_{t_0}^t dt' \; v(t')$.
\par
In summary, we have proposed a novel queueing theoretical model, which facilitates to simulate
the departure rates $Q_i^{\rm dep}(t)$ of road sections $i$ and the 
average travel times $T_i(t)$ of vehicles in an efficient way.
Assuming a fundamental diagram with linear free and congested branches
(i.e. constant propagation velocities $V_i^0$ and $-c$ in free and congested traffic, respectively)
allowed to eliminate the traffic dynamics within homogeneous road sections apart from the location
of shock fronts, i.e. the moving interfaces between free and congested traffic.
By determining the travel times from the arrival and departure rates of vehicles, 
the model considerably differs from classical traffic models, in which they are determined 
from the spatio-temporal vehicle speeds. However, note that there is some relationship with 
cell-based approaches \cite{cell}, if we subdivide our large homogeneous road sections into smaller subsections (cells).
\par
Apart from numerical efficiency, the
proposed section-based model is promising for analytical investigations of traffic in road networks
and of dynamic assignment problems. Moreover, it can describe the hysteretic breakdown of traffic
flow and reproduce typical congestion patterns. Nevertheless, it is not as accurate as
other macroscopic traffic models \cite{macro}, which do not assume constant propagation velocities
and can describe emergent stop-and-go waves. Despite of this, the queueing-theoretical 
traffic model is expected to provide reasonable estimates of the average travel times.
\par
The above model can be extended in several ways: (i) While the model can describe boundary-induced 
stop-and-go waves (characterized by the temporary disappearance of a
bottleneck $I_i Q_{\rm out}(t) - Q_i^{\rm cap}(t)$), emergent stop-and-go waves 
require the formulation of a generalized model containing a dynamic velocity equation,
which will reduce analytical tractability.
(ii) Accidents may be treated by a splitting of sections at the location and for the duration of an accident.
(iii) Stochastic effects in traffic flows could be easily incorporated.
(iv) Forecasts of travel times would be possible,
if the above model would be combined with a model for the prediction of origin-destination flows.
In such kinds of simulations, it makes sense to extend the model to multi-destination flows \cite{cell}. 
(v) Finally, the above model may
be generalized to the treatment of urban road networks with traffic lights, if we specify the
capacities $Q_i^{\rm cap}(t)$ in a time-dependent way which reflects the effects of amber and red lights.
Note that the actual section capacity does not change immediately with the traffic light,
but with some delay due to reaction and acceleration or deceleration times.
\par
{\em Acknowledgments:} The author is grateful to Martin Treiber, Martin Sch\"onhof, Thomas Seidel,
and an anonymous referee for useful comments.

\end{document}